\documentclass[prd,twocolumn,floats,floatfix,nofootinbib,superscriptaddress]{revtex4}
\usepackage{graphicx}
\usepackage{graphics}
\usepackage{dcolumn}
\usepackage{amssymb}
\usepackage{bm}
\usepackage{amsmath}
\def\spose#1{\hbox to 0pt{#1\hss}}

\def\lta{\mathrel{\spose{\lower 3pt\hbox{$\mathchar"218$}}
     \raise 2.0pt\hbox{$\mathchar"13C$}}}
\def\gta{\mathrel{\spose{\lower 3pt\hbox{$\mathchar"218$}}
     \raise 2.0pt\hbox{$\mathchar"13E$}}}
\newcommand{\ov}[1]{\overleftrightarrow{#1}}
\newcommand{\be}{\begin{equation}}
\newcommand{\en}{\end{equation}}
\newcommand{\bea}{\begin{eqnarray}}
\newcommand{\ena}{\end{eqnarray}}

\def\setR{\mathbb{R}}

\newcommand{\ie}{\textsl{i.e.~}}

\newcommand{\dee}{\mathrel{\mathop:}=}
\newcommand{\no}{\nonumber\\}
\def\1{1 \!\! 1}
\newcommand{\Ref}[1]{(\ref{#1})}
\def\ra{\rangle}
\def\la{\langle}
\newcommand{\ket}[1]{|{#1}\ra}

\newcommand{\bra}[1]{\la {#1}|}

\begin{document}
\title{The Wheeler-DeWitt Quantization Can Solve the Singularity Problem}

\author{F.~T.~Falciano}
\affiliation{ICRA - Centro Brasileiro de
Pesquisas F\'{\i}sicas -- CBPF, Rua Xavier Sigaud, 150, Urca,
CEP22290-180, Rio de Janeiro, Brazil}

\author{Roberto Pereira}
\affiliation{ICRA - Centro Brasileiro de
Pesquisas F\'{\i}sicas -- CBPF, Rua Xavier Sigaud, 150, Urca,
CEP22290-180, Rio de Janeiro, Brazil}

\author{N.~Pinto-Neto}
\affiliation{ICRA - Centro Brasileiro de
Pesquisas F\'{\i}sicas -- CBPF, Rua Xavier Sigaud, 150, Urca,
CEP22290-180, Rio de Janeiro, Brazil}

\author{E.~Sergio Santini}
\affiliation{ICRA - Centro Brasileiro de
Pesquisas F\'{\i}sicas -- CBPF, Rua Xavier Sigaud, 150, Urca,
CEP22290-180, Rio de Janeiro, Brazil}
\affiliation{CNEN - Comiss\~ao Nacional de Energia Nuclear \\
Rua General Severiano 90, Botafogo 22290-901,  Rio de Janeiro, Brazil}

\date{\today}

\begin{abstract}
We study the Wheeler-DeWitt quantum cosmology of a spatially flat Friedmann cosmological model with a massless free scalar field. We compare the consistent histories approach with the de~Broglie-Bohm theory when applied to this simple model under two different quantization schemes: the Schr\"odinger-like quantization, which essentially takes the square-root of the resulting Klein-Gordon equation through the restriction to positive frequencies and their associated Newton-Wigner states, or the induced Klein-Gordon quantization, that allows both positive and negative frequencies together. We show that the consistent histories approach can give a precise answer to the question concerning the existence of a quantum bounce if and only if one takes the single frequency approach and within a single family of histories, namely, a family containing histories concerning properties of the quantum system at only two specific moments of time: the infinity past and the infinity future. In that case, as shown by Craig and Singh \cite{CS}, there is no quantum bounce. In any other situation, the question concerning the existence of a quantum bounce has no meaning in the consistent histories approach. On the contrary, we show that if one considers the de~Broglie-Bohm theory, there are always states where quantum bounces occur in both quantization schemes. Hence the assertion that the Wheeler-DeWitt quantization does not solve the singularity problem in cosmology is not precise. To address this question, one must specify not only the quantum interpretation adopted but also the quantization scheme chosen.
\end{abstract}


\maketitle

\section{Introduction}\label{intro}

It has been claimed in many papers (some few examples are Refs.~\cite{CS,ash1,ash2,ash3,ash4}) that the Wheeler-DeWitt approach to quantum cosmology \cite{deWitt,halli3} does not solve the singularity problem of classical cosmology. This quite strong and general assertion can be criticized in many ways. First of all, this claim is usually based on calculations on a very simple model, namely, a free massless scalar field in Friedmann models, which is of course a very narrow subset of cosmological models. Secondly, the quantization program which is carried out on those papers is very particular: the Wheeler-DeWitt equation of these models are Klein-Gordon like equation, and the procedure is to extract a square root of it and work in a single frequency sector. Note that this is not mandatory, and there are other ways to deal with the Klein-Gordon equation working in the two frequency sectors with a well defined inner product, as it can be seen in Refs.~\cite{halli,halli2}. Furthermore, inside this single frequency sectors, the analysis is a priori restricted to quantum states which are either right-moving (expanding classical solutions) or left-moving (contracting classical solutions) \cite{ash1,ash2,ash3,ash4}. Finally, in some cases, the interpretation of the quantum states is not described clearly \cite{ash1,ash2,ash3}. Expectation values of the volume operator are calculated, but what do they mean? Of course they are not averages of external observers measurements of the volume operator as long as we are dealing with cosmology. Hence, are these expectation values interpreted along the lines of the many worlds interpretation \cite{everett}, the consistent histories approach \cite{hartle}, or something else? Among the above mentioned papers, only Refs.~\cite{ash4,CS} identify which interpretation is being used.

In the framework of a single frequency quantization of the simple model presented above, reference \cite{CS} presents the most rigorous and precise approach to the question of the existence of a quantum bounce in the Wheeler-DeWitt quantization. The interpretation they adopt is precisely defined (the consistent histories approach), and the conclusion is that, if one takes family of histories with properties defined in just two moments of time, the infinity past and the infinity future, then the probability of a quantum bounce is null for any quantum state, including superpositions of right and left-moving states. This is a remarkable result.

The aim of this paper is to discuss the results of Ref.~\cite{CS} with care, and contextualize it in the framework of other interpretations of quantum mechanics, namely, the de~Broglie-Bohm theory, and other quantization techniques, as the two frequencies (Klein-Gordon) approach of Ref.~\cite{halli}.

We will first show that, in the single frequency approach using the consistent histories interpretation, families of histories containing properties defined in one or more moments of time, besides properties defined in the infinity past and in the infinity future, are no longer consistent, unless one takes semi-classical states, which of course corresponds to histories without a bounce. This means that in the framework of these families of histories one cannot answer whether quantum bounces take place because histories involving any genuine quantum states are inconsistent. Hence the consistent histories approach is silent about quantum bounces happening in family of histories with more than two moments of time. Furthermore, we will show that in the induced Klein-Gordon approach there are no consistent family of histories involving genuine quantum states. Again, the question about the existence of quantum bounces has no meaning in the induced Klein-Gordon approach.

On the contrary, if one considers the de~Broglie-Bohm theory, where trajectories in configuration space are considered to be objectively real (the so called Bohmian trajectories), there is a vast number of examples of non-singular models. In this case, one can show that in the two quantization procedures mentioned above, there exist plenty of bouncing trajectories which tend to the classical cosmological trajectories when the volume of the universe is big. Similar results have already been obtained in earlier works (see Refs.~\cite{nelson1}-\cite{nelson2}). Hence, the existence of quantum bounces in the Wheeler-DeWitt approach depends strongly on the quantum interpretation one is adopting, and on the quantization procedure one is taking.

The paper will be divided as follows: in section II we will present the minisuperspace model we will work on, and summarize the results of Ref.~\cite{CS}. In section III, still in the framework of Ref.~\cite{CS}, we will consider families of histories with properties defined in three or more moments of time, and we will show that they are not consistent for general quantum states. In section IV we will summarize the de~Broglie-Bohm theory applied to quantum cosmology and we will present an infinitude of Bohmian trajectories which are non-singular and approach the classical limit for large volumes of space. In section V we show that there are no families of consistent histories in the Klein-Gordon quantization for general quantum states. However, also in this quantization scheme, there are many Bohmian trajectories which are non-singular. We end up with the conclusions in section VI.

\section{The Craig and Singh result}\label{CSsec}

In this section we shall briefly develop the Wheeler-DeWitt quantization of a FLRW model with flat spatial sections following closely reference \cite{CS}. We assume that the matter content of the universe can be described by a massless scalar field and we consider the consistent histories approach of \cite{hartle} to explore the existence of a quantum bounce. The Hamiltonian constraint of this minisuperspace system reads
\be
H=  -\frac{2\pi G}{3}\frac{p_a^2}{a} + \frac{p_\phi^2}{2a^3} \approx 0,
\en
where $a$ is the scale factor, $\phi$ the scalar field and $p_a,p_\phi$ their conjugate momenta, given by
\be
p_a= - \frac{3}{4\pi G} a\dot{a} \;\; , \;\; p_\phi=a^3\dot{\phi}.
\en

On the derivation of the Hamiltonian above the lapse function was set to one and the fiducial cell considered to have a unit volume. Denoting $\alpha\dee \log a$, we rewrite the Hamiltonian in a more convenient form
\be
H= \frac{e^{-3\alpha}}{2}\left(-\frac{4\pi G}{3} p_\alpha^2 + p_\phi^2 \right) \approx 0.
\en

Note that our choice of variables is different of the one used in \cite{CS}. The conclusions are nevertheless unchanged. The pair $(\alpha. p_\alpha)$ is still canonical. After quantization the momenta promoted to derivative operators and the classical constraint $H\approx 0$ becomes the Wheeler-deWitt equation
\be
\left(\partial_\phi^2 - \frac{4\pi G}{3} \partial_\alpha^2\right) \Psi(\alpha,\phi) = 0,
\en
defined on the kinematical Hilbert space $L^2(\setR^2,d\alpha d\phi)$. We see that the Wheeler-deWitt quantization of this simple model is equivalent to the quantization of the Klein-Gordon equation in Minkowski spacetime. The standard procedure is to separate the positive and negative frequency modes and quantize them independently. Taking the square-root of the constraint, we get
\be
\pm i\partial_\phi \Psi(\alpha.\phi) = \sqrt{\mathbf{\Theta}} \ \Psi(\alpha,\phi),
\en
with
\be \mathbf{\Theta} \dee -\frac{4\pi G}{3} \partial_\alpha^2.\en

The action of $\sqrt{\mathbf{\Theta}}$ is best seen on Fourier space. Consider the set of eigenfunctions
\be\label{eigenfunc}
e_k(\alpha) = \bra{\alpha} k\ra = \frac{1}{\sqrt{2\pi}} e^{ik\alpha},
\en
such that $\mathbf{\Theta} \ e_k = \omega^2 e_k$, with
\be
\omega\dee \sqrt{\frac{4\pi G}{3}} |k|.
\en

The physical scalar product is given by
\be
\bra{\Phi}\Psi\ra \dee \int_{\phi=\phi_0} d\alpha\,  \bar{\Phi}(\alpha,\phi) \Psi(\alpha,\phi),
\en
and it is independent of the time $\phi_0$ on which it is defined. Positive and negative frequency sectors are orthogonal with respect to this scalar product. Restricting to the positive frequency sector, evolution is given by the propagator
\be
U(\phi-\phi_0)= e^{i\sqrt{\mathbf{\Theta}} (\phi-\phi_0)}
\en

There are two Dirac observables. The first one is $p_\phi$, which is an invariant of the model. The second one is a relational observable \cite{relational}. For any operator $\hat{A}$, which does not need to commute with the Hamiltonian, one constructs the corresponding relational observable
\be
\hat{A}|_{\phi_0}\ket{\Psi(\phi)} = U(\phi-\phi_0) \hat{A}\ket{\Psi(\phi_0)},
\en
which implies
\be
\hat{A}|_{\phi_0}=U(\phi_0-\phi)^\dag \hat{A} U(\phi_0-\phi).
\en
Applying this to the operator $\hat{\alpha}$, the action of the Dirac observables on the positive frequency sector is given by
\bea
&&\hat{\alpha}|_{\phi_0}  \Psi(\alpha,\phi) = U(\phi-\phi_0) \hat{\alpha}  \Psi(\alpha,\phi_0) \no
&& \hat{p}_\phi   \Psi(\alpha,\phi) = \hbar\sqrt{\mathbf{\Theta}}   \Psi(\alpha,\phi).
\ena
Note that the action of the Dirac observables preserves the positive and negative subspaces, which is consistent with the approach taken.

With the physical Hilbert space specified, we need to construct now the set of histories and the corresponding decoherence functional.

Following Hartle's approach \cite{hartle} we are interested in defining a decoherence functional for a set of histories. The decoherence functional is defined as
\be
d(h,h')\dee \bra{\Psi_{h'}}\Psi_h\ra,
\en
where the branch wave function is given by
\be
\Psi_h\dee C_h^\dag\ket{\Psi}.
\en
In the above formula, $\Psi$ is a given initial state and $C_h$ is the class operator defining the history $h$, given by a product of projectors
\be
C_h\dee P^{\mathcal{O}_1}_{\Delta\lambda_{k_1}}(t_1) ... P^{\mathcal{O}_n}_{\Delta\lambda_{k_n}}(t_n).
\en
$P^{\mathcal{O}}_{\Delta\lambda_{k}}(t)$ projects onto the subspace for which the $k$th eingenvalue of the observable $\mathcal{O}$ at time $t$ takes values in the interval $\Delta\lambda_{k}$. Here we are using Heisenberg operators for the projectors
\be
P^{\mathcal{O}}_{\Delta\lambda_{k}}(t) \dee U^\dag(t) P^{\mathcal{O}}_{\Delta\lambda_{k}} U(t).
\en

For the case at hand we consider the observable given by the scale factor $a$, or $\alpha$, with relational time $\phi$, and we will denote projectors simply by $P_{\Delta\alpha_i}(\phi_i)$. The time independent projector is given explicitly by
\be
P_{\Delta\alpha} = \int_{\Delta\alpha} d\alpha \ket{\alpha}\bra{\alpha},
\en
where the ket $\ket{\alpha}$ is defined in \Ref{eigenfunc} and the normalization is such as to make this basis orthonormal. Let us further compute the expectation value of the evolution operator in this basis, as it will be useful in the next section
\be\label{defU}
\bra{\alpha'} U(\phi) \ket{\alpha} = \int \frac{dk}{2\pi} e^{i \omega \phi} e^{ik(\alpha'-\alpha)},
\en
where we used the resolution of the identity on Fourier space
\be
\int_\setR dk \ket{k}\bra{k} = \1.
\en

The set of histories considered in \cite{CS} are composed by two times, corresponding the past infinity ($\phi\rightarrow -\infty$) and future infinity ($\phi\rightarrow +\infty$). The histories are separated in those where $\alpha$ is bigger or smaller than a given fixed fiducial value $\alpha_*$. There are four possible histories according to these possibilities, described by the following class operators
\bea
C_{S-S}(-\infty,\infty) &=& P_{\Delta\alpha_1}(-\infty) P_{\Delta\alpha_2}(\infty) \no
C_{S-B}(-\infty,\infty) &=& P_{\Delta\alpha_1}(-\infty) P_{\bar{\Delta}{\alpha_2}}(\infty) \no
C_{B-S}(-\infty,\infty) &=& P_{\bar{\Delta}{\alpha_1}}(-\infty) P_{\Delta\alpha_2}(\infty) \no
C_{B-B}(-\infty,\infty) &=& P_{\bar{\Delta}{\alpha_1}}(-\infty) P_{\bar{\Delta}{\alpha_2}}(\infty) \nonumber
\ena
where $S$ and $B$ subscripts denote, respectively, the domains of the scale factor arbitrarily close to the singularity
or arbitrarily big.

Let us now check that this set of histories is consistent. Noting that $P_{\Delta\alpha} P_{\bar{\Delta}{\alpha}} = 0$, the only non-trivial terms are $d(h_{S-B},h_{B-B})$ and $d(h_{S-S},h_{B-S})$. Consider for example the first of these terms
\be\label{dfunc}
d(h_{S-B},h_{B-B}) = \bra{\Psi} P_{\bar{\Delta}{\alpha_1}}(\phi_1) P_{\bar{\Delta}{\alpha_2}}(\phi_2) P_{\Delta\alpha_1}(\phi_1) \ket{\Psi}.
\en
Let us as a first step study the behavior of $P_{\Delta\alpha}(\phi) \ket{\Psi}$ and $P_{\bar{\Delta}{\alpha}}(\phi)$ for $\phi\rightarrow\pm\infty$. We borrow the results without proof from \cite{CS}. We have that
\bea
&&\lim_{\phi\rightarrow +\infty} P_{\Delta\alpha}(\phi) \ket{\Psi} = \ket{\Psi_L} \no
&&\lim_{\phi\rightarrow -\infty} P_{\Delta\alpha}(\phi) \ket{\Psi} = \ket{\Psi_R} \no
&&\lim_{\phi\rightarrow +\infty} P_{\bar{\Delta}{\alpha}}(\phi) \ket{\Psi} = \ket{\Psi_R} \no
&&\lim_{\phi\rightarrow -\infty} P_{\bar{\Delta}{\alpha}}(\phi) \ket{\Psi} = \ket{\Psi_L}. \no
\ena

In the equation above we have used the left/right-moving decomposition of the wave function
\bea
\Psi(\alpha,\phi)&&= \frac{1}{\sqrt{2\pi}} \int_\setR dk \, \Psi(k) e^{ik\alpha} e^{i\omega \phi} \no
&\propto& \int_{-\infty}^0 dk \, \Psi(k)e^{ik(\alpha - \phi)}+ \int_0^\infty dk \, \Psi(k)e^{ik(\alpha + \phi)}  = \no
&=& \Psi_R(v_r) + \Psi_L(v_l),
\ena
where $v_r \dee \alpha - \phi$, $v_l\dee \alpha+\phi$ and we dropped the factor of $\sqrt{4\pi G/3}$ in $\omega$.

Since right and left moving sectors are orthogonal, the term in \Ref{dfunc} is zero, as is the other term, and the decoherence functional is diagonal for this set of histories.

Craig and Singh go on and restrict the histories to those giving rise to a bounce and those that have a singularity and prove that the probability for a bounce is zero.

As noted by the authors in \cite{CS}, the decoherence functional in the case with more than two times is not necessarily diagonal, unless the wave functions are taken to be semiclassical. In the next section we will show explicitly that already in the case of three times the decoherence functional is not diagonal.

\section{Histories with three times}\label{3hist}

Let us now see whether we can obtain a family of consistent histories when we ask about properties concerning the size of the universe in a third moment of time between $\phi_1 \to -\infty$ and $\phi_2\to +\infty$. Thus we want to address the question whether in an arbritary intermediary $\phi$ time
the scale factor of the universe is in the interval $(-\infty, \alpha *)$, or on its complement $(\alpha *,\infty)$.

The new family has now eight histories associated with the following class operators

\begin{eqnarray}
\label{classS2}
C_{S-\Delta\alpha -S}(\phi_1,\phi,\phi_2)&=&P_{\Delta\alpha_1}(\phi_1)P_{\Delta\alpha}(\phi)
P_{\Delta\alpha_2}(\phi_2),\nonumber\\
C_{S-\bar{\Delta}\alpha-S}(\phi_1,\phi,\phi_2)&=&P_{\Delta\alpha_1}(\phi_1)P_{\bar{\Delta}\alpha}(\phi)
P_{\Delta\alpha_2}(\phi_2),\nonumber\\
C_{S-\Delta\alpha -B}(\phi_1,\phi,\phi_2)&=&P_{\Delta\alpha_1}(\phi_1)P_{\Delta\alpha}(\phi)
P_{\bar{\Delta}\alpha_2}(\phi_2),\nonumber\\
C_{S-\bar{\Delta}\alpha -B}(\phi_1,\phi,\phi_2)&=&P_{\Delta\alpha_1}(\phi_1)P_{\bar{\Delta}\alpha}(\phi)
P_{\bar{\Delta}\alpha_2}(\phi_2),\nonumber\\
C_{B-\Delta\alpha -S}(\phi_1,\phi,\phi_2)&=&P_{\bar{\Delta}\alpha_1}(\phi_1)P_{\Delta\alpha}(\phi)
P_{\Delta\alpha_2}(\phi_2),\nonumber\\
C_{B-\bar{\Delta}\alpha -S}(\phi_1,\phi,\phi_2)&=&P_{\bar{\Delta}\alpha_1}(\phi_1)P_{\bar{\Delta}\alpha}(\phi)
P_{\Delta\alpha_2}(\phi_2),\nonumber\\
C_{B-\Delta\alpha -B}(\phi_1,\phi,\phi_2)&=&P_{\bar{\Delta}\alpha_1}(\phi_1)P_{\Delta\alpha}(\phi)
P_{\bar{\Delta}\alpha_2}(\phi_2),\nonumber\\
C_{B-\bar{\Delta}\alpha -B}(\phi_1,\phi,\phi_2)&=&P_{\bar{\Delta}\alpha_1}(\phi_1)P_{\bar{\Delta}\alpha}(\phi)
P_{\bar{\Delta}\alpha_2}(\phi_2)
\end{eqnarray}

with $S$ and $B$ having the same meaning as before being close to the singularity or arbitrarily big.

Each of these class operators is associated with a particular
history. For instance, the class operator $C_{S-\Delta\alpha -B}(\phi_1,\phi,\phi_2)$ is associated
with the history where the universe was singular at $\phi_1\to -\infty$, has a size in a domain $\Delta\alpha$
at the finite time $\phi$, and it will be infinitely large at $\phi_2\to\infty$.

One must now see whether this new family with eight histories is consistent or not. As we have seen above, one must calculate the decoherence functional $d(h,h')$ for the histories associated with the class operators shown in Eq.~(\ref{classS2}).

It is easy to show that, in general, $d(h,h')$ is not approximately zero. For that, let us calculate the decoherence functional for the histories associated with the class operators $C_{S-\Delta\alpha -B}(\phi_1,\phi,\phi_2)$ and $C_{S-\bar{\Delta}\alpha -B}(\phi_1,\phi,\phi_2)$. In this case the functional reads
\begin{widetext}

\begin{eqnarray}
\label{dSc}
d(h_{S-\bar{\Delta}\alpha -B} , h_{S-\Delta\alpha -B}) &=&
{\rm Tr} (P_{\bar{\Delta}\alpha_2}(\phi_2)P_{\bar{\Delta}\alpha}(\phi)
P_{\Delta\alpha_1}(\phi_1)\ket{\Psi}\bra{\Psi}P_{\Delta\alpha_1}(\phi_1)P_{\Delta\alpha}(\phi)
P_{\bar{\Delta}\alpha_2}(\phi_2)) ,\nonumber\\
&=&\bra{\Psi_R}P_{\Delta\alpha}(\phi)P_{\bar{\Delta}\alpha_2}P_{\bar{\Delta}\alpha}(\phi)\ket{\Psi_R} ,
\end{eqnarray}
where we have used that $P_{\Delta\alpha_1}(\phi_1\to -\infty)\ket{\Psi} = \ket{\Psi_R}$. Then, dropping constant factors in front of the integrals and using eq. \Ref{defU}, we have
\begin{align}
\label{dSc2}
d(h_{S-\Delta\alpha -B},h_{S-\bar{\Delta}\alpha -B}) =
\lim _{\phi_2\to\infty} \int_{-\infty}^{\alpha *}d\alpha '' \Psi^{*} (v_r '')
&\int_{\alpha *}^{\infty} d\alpha \Psi (v_r) \int_{\alpha_2 *}^{\infty} d\alpha '
<\alpha ''|U(\phi - \phi_2)|\alpha '><\alpha '|U(\phi_2 - \phi)|\alpha> \nonumber \\
=\lim _{\phi_2\to\infty} \int_{-\infty}^{\alpha *}d\alpha '' \Psi^{*} (v_r '')
\int_{\alpha *}^{\infty} d\alpha \Psi (v_r) \int_{\alpha_2 *}^{\infty} d\alpha '&
\biggl(\int_{0}^{\infty} dk'  \ e^{ -i k ' (\alpha ' - \alpha ''+\phi_2 -\phi)}+
\int_{-\infty}^{0} dk ' \ e^{-i k ' (\alpha ' - \alpha ''+\phi -\phi_ 2)}\biggr)\times \nonumber \\
& \biggl(\int_{0}^{\infty} dk  \ e^{-i k  (\alpha  - \alpha '+\phi -\phi_2)}+
\int_{-\infty}^{0} dk  \ e^{-i k  (\alpha  - \alpha '+\phi_2 -\phi)}\biggr)\quad.
\end{align}

We shall analyse each of its four terms separately. The term formed by the product of the first with the third can be written as
\begin{equation}
\label{dec11}
\lim _{\phi_2\to\infty} \int_{\alpha_2 * + \phi_2}^{\infty} \ e^{i v_l ' (k  - k')} dv_l '
\int_{\alpha *}^{\infty} \Psi (v_r) d\alpha \int_{-\infty}^{\alpha *}d\alpha '' \Psi^{*} (v_r '')
\int_{0}^{\infty} dk'  \ e^{i k ' v_l ''}
\int_{0}^{\infty} dk  \ e^{-i k  v_l} =0\quad ,
\end{equation}
while the product of the second with the fourth gives
\begin{eqnarray}
\label{dec22}
&&\lim _{\phi_2\to\infty} \int_{\alpha_2 * - \phi_2}^{\infty} \ e^{i v_r ' (k - k')} dv_r '
\int_{\alpha *}^{\infty} \Psi (v_r) d\alpha \int_{-\infty}^{\alpha *}d\alpha '' \Psi^{*} (v_r '')
\int_{-\infty}^{0} dk'  \ e^{i k ' v_r ''}
\int_{-\infty}^{0} dk  \ e^{-i k  v_r} \nonumber \\
&=& \int_{\alpha *}^{\infty} \Psi (v_r) d\alpha \int_{-\infty}^{\alpha *}d\alpha '' \Psi^{*} (v_r '')
\int_{0}^{\infty} dk \ e^{-i k  (v_r '' - v_r)} \nonumber \\
&=& \int_{\alpha * - \phi}^{\infty} \Psi (v_r) dv_r \int_{-\infty}^{\alpha * - \phi}dv_r '' \Psi^{*} (v_r '')
\biggl[\pi\delta (v_r '' - v_r) + i{\rm p.v.}\biggl(\frac{1}{v_r '' - v_r}\biggr)\biggr]\nonumber\\
&=&  i{\rm p.v.}\int_{\alpha * - \phi}^{\infty} dv_r \int_{-\infty}^{\alpha * - \phi} dv_r ''
\biggl(\frac{\Psi (v_r)\Psi^{*} (v_r '')}{v_r '' - v_r}\biggr),
\end{eqnarray}
where in the last equality we have used the fact that the integral in $v_r$ and $v_r ''$ are in disjoint domains, and hence the part involving the delta function is null. For the definition of the principal value, noted above as p.v., see appendix \ref{appdist}. The sum of the cross terms involving the products of the first with fourth terms, and the second and third reads, respectively,
\begin{eqnarray}
\label{dec1221}
&&\lim _{\phi_2\to\infty}
\int_{\alpha *}^{\infty} \Psi (v_r) d\alpha \int_{-\infty}^{\alpha *}d\alpha '' \Psi^{*} (v_r '')
\int_{\alpha_2 * - \phi_2}^{\infty} dv_r ' \biggl(\int_0^{\infty} dk'  \ e^{-i k ' \left[v_r ' - v_r '' + 2(\phi_2 - \phi)\right]}
\int_0^{\infty} dk  \ e^{-i k  (v_r ' - v_r)} \nonumber \\ &+&
\int_0^{\infty} dk  \ e^{-i k  \left[v_r  - v_r ' - 2(\phi_2 - \phi)\right]}
\int_0^{\infty} dk '  \ e^{-i k '(v_r '' - v_r ')} \biggr) .
\end{eqnarray}
Let us concentrate on the first term. The other term follows the same reasoning. Performing the integrals on $k$ and $k '$, one gets
\begin{eqnarray}
\label{dec12}
&&\lim _{\phi_2\to\infty}
\int_{\alpha * -\phi}^{\infty} \Psi (v_r) dv_r \int_{-\infty}^{\alpha * -\phi}dv_r '' \Psi^{*} (v_r '')
\int_{\alpha_2 * - \phi_2}^{\infty} dv_r ' \nonumber \\
&&\left[ \pi\delta (v_r ' - v_r '' + 2(\phi_2 - \phi)) - i{\rm p.v.}\biggl(\frac{1}{v_r ' - v_r '' +
2(\phi_2 - \phi)}\biggr)\right]
\left[ \pi\delta (v_r ' - v_r) - i{\rm p.v.}\biggl(\frac{1}{v_r ' - v_r}\biggr)\right]\quad.
\end{eqnarray}
The typical terms are proportional to
\begin{eqnarray}
\label{dec12b}
&&\lim _{\phi_2\to\infty}
\int_{-\infty}^{\alpha * -\phi}dv_r '' \Psi^{*} (v_r '')\Psi (v_r '' - 2(\phi_2 - \phi)) \quad,\nonumber \\
&& \lim _{\phi_2\to\infty}
 {\rm p.v.}\int_{-\infty}^{\alpha * -\phi}dv_r '' \Psi^{*} (v_r '')
\int_{\alpha * -\phi}^{\infty} dv_r \biggl(\frac{\Psi (v_r) }{v_r '' - v_r -
2(\phi_2 - \phi)}\biggr)\quad,\nonumber \\
&& \lim _{\phi_2\to\infty}
 {\rm p.v}\int_{-\infty}^{\alpha * -\phi}dv_r '' \Psi^{*} (v_r '')
\int_{\alpha * -\phi}^{\infty} dv_r \Psi (v_r)\int_{\alpha_2 *}^{\infty} d\alpha '
\biggl(\frac{1}{(\alpha '- v_r '' +
\phi_2 - 2\phi)(\alpha '- v_r - \phi_2)}\biggr),
\end{eqnarray}
The terms with $\Psi^{*} (v_r '' - 2(\phi_2 - \phi))$ are zero in the limit $\phi_2\to\infty$ because
$\Psi$ is assumed to be square integrable, while the terms with a $\phi_2$ in denominators
are obviously null in this limit.

The only non-null term of Eq.~(\ref{dSc2}) is Eq.~(\ref{dec22}). Hence, the final result for this off-diagonal term of the decoherence functional is

\begin{equation}
\label{decfinal}
d(h_{S-\bar{\Delta}\alpha -B},h_{S-\Delta\alpha -B}) \propto - i{\rm p.v.}\int_{\alpha * - \phi}^{\infty} dv_r \int_{-\infty}^{\alpha * - \phi} dv_r ''\biggl[\frac{\Psi (v_r)\Psi^{*} (v_r '')}{v_r '' - v_r}\biggr],
\end{equation}
which is not null in general. Due to the disjoint domains of integration, this result can be approximately zero if and only if $\Psi (v_r)$ is concentrated around some fixed value of $v_r$. The classical trajectories are given by $v_r =$ const or $v_l =$ const. Therefore, a wave function sharply concentrated around some fixed value of $v_r$ must describe a semiclassical state. It is straightforward to show that other off-diagonal terms of the decoherence functional,
e.g., $d(h_{B-\Delta\alpha -S},h_{B-\bar{\Delta}\alpha -S})$, are approximately zero only if the wave function $\Psi (v_l)$ is concentrated around the other class of classical trajectories $v_l =$ const.

\end{widetext}

Concluding, the family of histories described by the class operators (\ref{classS2}) can be made consistent only
for semiclassical states. In that case, of course, the probability of occurrence of a quantum bounce
is null, as before, but the reason for that comes from the fact that we are not allowed to calculate
probabilities in a family of cosmological histories where quantum effects are relevant. Probabilities
are calculable only for semiclassical histories, where bounces cannot occur.
More generally, if quantum effects are important in any family of cosmological histories,
under the consistent histories approach one cannot ask any questions about properties of the
Universe at an arbitrary finite $\phi$. This is of course a limitation on the applicability of the
consistent histories approach to cosmology, at least for the present simple model. We are simply
prohibited to study the quantum properties of a cosmological model, unless one considers just
two moments of its history, at $\phi \pm \infty$, and nothing more than that. Are there any other
approaches to quantum cosmology where one can go further?

\section{The de~Broglie-Bohm theory applied to quantum cosmology}\label{dBBqc}

A quantum theory that can be consistently implemented in the quantum cosmology scenario is the de~Broglie-Bohm quantum theory (see Ref.'s \cite{dBB}-\cite{kowalski2} for details). Considering minisuperspace models, which have a finite number
of degrees of freedom, the general form of the associated Wheeler-De Witt equation reads
\begin{equation}
\label{bsc}
-\frac{1}{2}f_{\rho\sigma}(q_{\mu})\frac{\partial \Psi (q)}{\partial q_{\rho}\partial q_{\sigma}}
+ U(q_{\mu})\Psi (q) = 0 \quad,
\end{equation}
where $f_{\rho\sigma}(q_{\mu})$ is the minisuperspace DeWitt metric of the model, whose inverse is denoted by $f^{\rho\sigma}(q_{\mu})$. By writing the wave function in its polar form, $\Psi = R \ e^{iS}$, the complex equation (\ref{bsc}) decouples in two real equations
\begin{equation}
\label{hoqg}
\frac{1}{2}f_{\rho\sigma}(q_{\mu})\frac{\partial S}{\partial q_{\rho}}
\frac{\partial S}{\partial q_{\sigma}}+ U(q_{\mu}) + Q(q_{\mu}) = 0 \quad,
\end{equation}
\begin{equation}
\label{hoqg2}
f_{\rho\sigma}(q_{\mu})\frac{\partial}{\partial q_{\rho}}
\biggl(R^2\frac{\partial S}{\partial q_{\sigma}}\biggr) = 0 \quad,
\end{equation}
where
\begin{equation}
\label{hqgqp}
Q(q_{\mu}) \dee -\frac{1}{2R} f_{\rho\sigma}\frac{\partial ^2 R}
{\partial q_{\rho} \partial q_{\sigma}}
\end{equation}
is called the quantum potential. The de~Broglie-Bohm interpretation applied to Quantum Cosmology states that the trajectories $q_{\mu}(t)$ are real, independently of any observations. Equation (\ref{hoqg}) represents their Hamilton-Jacobi equation, which is the classical one  added with a quantum potential term Eq.(\ref{hqgqp}) responsible for the quantum effects. This suggests to define

\begin{equation}
\label{h}
p^{\rho} = \frac{\partial S}{\partial q_{\rho}} ,
\end{equation}
where the momenta are related to the velocities in the usual way

\begin{equation}
\label{h2}
p^{\rho} = f^{\rho\sigma}\frac{1}{N}\frac{\partial q_{\sigma}}{\partial t} .
\end{equation}

In order to obtain the quantum trajectories, we have to solve the following
system of first order differential equations, called the guidance relations
\begin{equation}
\label{h3}
\frac{\partial S(q_{\rho})}{\partial q_{\rho}} =
f^{\rho\sigma}\frac{1}{N}\dot{q}_{\sigma} .
\end{equation}

The above equations (\ref{h3}) are invariant under time reparametrization. Therefore, even at the quantum level, different time gauge choices of $N(t)$ yield the same space-time geometry for a given non-classical solution $q_{\alpha}(t)$. Indeed, there is no problem of time in the de~Broglie-Bohm interpretation for
minisuperspace quantum cosmological models \cite{bola27}. However, this is no longer true when one considers the full superspace (see \cite{santini0,tese}). Notwithstanding, even with the problem of time in the superspace, the theory can be consistently formulated (see \cite{tese,cons}).

Let us then apply this interpretation to our minisuperspace model.
The Wheeler-DeWitt equation then reads
\begin{equation}
\label{wdw}
-\frac{\partial ^2\Psi}{\partial \alpha ^2} +  \frac{\partial
^2\Psi}{\partial \phi ^2} = 0 \quad.
\end{equation}
Comparing Eq.~(\ref{wdw}) with Eq.~(\ref{bsc}), we obtain,
from Eqs.~(\ref{hoqg}) and (\ref{hoqg2}),

\begin{equation}
\label{hoqgp}
- \biggl(\frac{\partial S}{\partial \alpha}\biggr)^2 + \biggl(\frac{\partial S}{\partial \phi}\biggr)^2
+ Q(q_{\mu}) = 0 \quad,
\end{equation}
\begin{equation}
\label{hoqg2p}
\frac{\partial}{\partial \phi} \biggl(R^2\frac{\partial S}{\partial \phi}\biggr)
- \frac{\partial}{\partial \alpha} \biggl(R^2\frac{\partial S}{\partial \alpha}\biggr) = 0 \quad,
\end{equation}
where the quantum potential reads
\begin{equation}
\label{qp1}
Q(\alpha ,\phi )\dee \frac{1}{R}\biggr[\frac{\partial^{2}R}
{\partial \alpha^{2}}-\frac{\partial^{2}R}{\partial \phi^{2}}\biggl]\quad .
\end{equation}
The guidance relations (\ref{h3}) are
\begin{equation}
\label{guialpha}
\frac{\partial S}{\partial \alpha}=-\frac{e^{3\alpha}\dot{\alpha}}{N}\quad ,
\end{equation}
\begin{equation}
\label{guiphi}
\frac{\partial S}{\partial \phi}=\frac{e^{3\alpha}\dot{\phi}}{N}\quad .
\end{equation}

We can write equation Eqs.~(\ref{hoqgp}) in null coordinates,
\begin{eqnarray}
\label{nulas}
v_l\dee\frac{1}{\sqrt{2}}(\alpha+\phi) \quad & &\quad  \alpha\dee\frac{1}{\sqrt{2}} \left(v_l+v_r\right)\nonumber\\
v_r\dee\frac{1}{\sqrt{2}}(\alpha-\phi) \quad  & &\quad  \phi\dee\frac{1}{\sqrt{2}} \left(v_l-v_r\right)
\end{eqnarray}
yielding,
\begin{equation}
\left(-\frac{\partial^{2} }{\partial v_l \partial v_r }\right) \Psi \left(v_l,v_r \right) =0 \quad.
\end{equation}
The general solution is
\begin{equation}
\label{sol0}
\Psi(u,v) = F(v_l) + G(v_r) \quad,
\end{equation}
where $F$ and $G$ are arbitrary functions.
Using a separation of variable method, one can write these solutions
as Fourier transforms given by
\begin{equation}
\label{sol0k}
\Psi(v_l,v_r) = \int_{-\infty}^{\infty} d k U(k)\ e^{ikv_l} + \int_{-\infty}^{\infty} d k V(k)\ e^{ikv_r} \quad,
\end{equation}
with $U$ and  $V$ also being two arbitrary functions. If one restricts the wave function
Eq.~(\ref{sol0k}) to left or right moving components only, the quantum potential
will necessarily be a function of either $v_l$ or $v_r$, and hence it will be null
(see Eq.~(\ref{qp1})). In this case, only classical trajectories, which are of course
singular, are allowed. Hence, avoidance of singularities is possible if and only
if the wave function Eq.~(\ref{sol0k}) depends on both left and right moving components.

Under restriction to positive frequency solutions, one gets a subclass of the general solution Eq.~(\ref{sol0k})

\begin{equation}
\label{solf+}
\Psi(v_l,v_r) = \int_{0}^{\infty} d k \Psi(k)\ e^{ikv_l} + \int_{-\infty}^{0} d k \Psi(k)\ e^{ikv_r} \quad .
\end{equation}
From the guidance equations (\ref{guialpha}) and (\ref{guiphi}), one obtains that

\begin{widetext}

\begin{eqnarray}
\label{guitotal}
\frac{d\alpha}{d\phi} &=& -\frac{\partial S/\partial \alpha}{\partial S /\partial \phi}\nonumber \\
&=& - \biggl\{\int_{0}^{\infty} dk  \int_{0}^{\infty} dk ' \biggl[\Psi (k) \Psi^{*} (k') \ e^{i v_l(k-k ')}
- \Psi (-k) \Psi^{*} (-k') \ e^{-i v_r(k-k ')}\biggr] (k+k ')\nonumber \\
&-& \biggl[\Psi (-k) \Psi^{*} (k') \ e^{-i v_r k}\ e^{-i v_l k '}
- \Psi (k) \Psi^{*} (-k') \ e^{i v_l k}]\ e^{i v_r k '}\biggr] (k-k ')\biggr\}/\nonumber \\
&& \biggl\{\int_{0}^{\infty} dk  \int_{0}^{\infty} dk ' \biggl[\Psi (k) \Psi^{*} (k') \ e^{i v_l(k-k ')}
+ \Psi (-k) \Psi^{*} (-k') \ e^{-i v_r(k-k ')} \nonumber \\
&+& \Psi (-k) \Psi^{*} (k') \ e^{-i v_r k}\ e^{-i v_l k '}
+ \Psi (k) \Psi^{*} (-k') \ e^{i v_l k}]\ e^{i v_r k '}\biggr] (k+k ')\biggr\} .
\end{eqnarray}

Let us analize Eq.~(\ref{guitotal}) in the limits $v_r\to\pm\infty$ or $v_l\to\pm\infty$.
When $v_r\to\pm\infty$, the integrals involving $\int_{0}^{\infty} dk \Psi (k) \ e^{i v_r k}$,
$\int_{0}^{\infty} dk k \Psi (k) \ e^{i v_r k}$ and similar correspond to a Fourier transform
of square integrable functions which are null when evaluated at $v_r\to\pm\infty$. Hence we obtain
from Eq.~(\ref{guitotal}), in this limit, that

\begin{equation}
\label{inftyr}
\frac{d\alpha}{d\phi} = -1 \ \Rightarrow \ \alpha + \phi = v_l = {\rm {const}}\quad.
\end{equation}

For $v_l\to\pm\infty$, an analogous reasoning yields

\begin{equation}
\label{inftyl}
\frac{d\alpha}{d\phi} = 1 \ \Rightarrow  \ \alpha - \phi = v_r = {\rm {const}}\quad.
\end{equation}

Hence, in the regions $v_r\to\pm\infty$ and $v_l\to\pm\infty$, the Bohmian trajectories emerging
from Eq.~(\ref{guitotal}) are the classical trajectories irrespectively of the wave function.

We will now see, however, that there are a huge
class of states where the Bohmian trajectories are not classical in other regions of the
$(\alpha, \phi)$ plane. For instance,
when $\Psi(k)$ is even on $k$, Eq.~(\ref{guitotal}) reads

\begin{equation}
\label{guitotalp}
\frac{d\alpha}{d\phi} = - i\frac{\int_{0}^{\infty} dk  \int_{0}^{\infty} dk ' \Psi (k) \Psi^{*} (k') \ e^{i \phi (k-k ')}
\{\sin [\alpha (k-k ')](k+k')+\sin [\alpha (k+k ')](k-k')\}}
{\int_{0}^{\infty} dk  \int_{0}^{\infty} dk ' \Psi (k) \Psi^{*} (k') \ e^{i \phi (k-k ')}
\{\cos [\alpha (k-k ')]+\cos [\alpha (k+k ')]\}(k+k')} .
\end{equation}

Note that Eq.~(\ref{guitotalp}) is anti-symmetric under the change $\alpha \rightarrow -\alpha$ and also $d\alpha / d\phi =0$ at $\alpha =0$. Consequently, the Bohmian trajectories that start at $v_r\to\infty$ (infinitely big universe) cannot cross the line $\alpha =0$ and goes to the singularity at $v_r\to -\infty$ in the same way as the classical trajectories do. These Bohmian trajectories are non-singular. On the other hand, if they start at the singularity in $v_l\to -\infty$, they cannot become infinitely big at $v_l\to \infty$.

Note that this result is in opposition to the consistent histories conclusion. These Bohmian trajectories describe exactly what the consistent histories approach has claimed to be impossible, namely, universe histories that start infinitely big in the far past and go infinitely big also in the far future. In fact, the even states shown above within the de~Broglie-Bohm scenario violate the consistent histories description for all trajectories. As just argued, there is no single trajectory that can start infinitely big in the far past and goes to a singularity in the far future or the reverse. Hence, it is certain that there exist non-singular
Bohmian trajectories.

One can also obtain bounces in the situation where $\Psi(k)$ is not only even on $k$ but it is also real. In that case
Eq.~(\ref{guitotalp}) simplifies to
\begin{equation}
\label{guitotalpr}
\frac{d\alpha}{d\phi} = \frac{\int_{0}^{\infty} dk  \int_{0}^{\infty} dk ' \Psi (k) \Psi^{*} (k') \sin [\phi (k-k ')]
\{\sin [\alpha (k-k ')](k+k')+\sin [\alpha (k+k ')](k-k')\}}
{\int_{0}^{\infty} dk  \int_{0}^{\infty} dk ' \Psi (k) \Psi^{*} (k') \cos [\phi (k-k ')]
\{\cos [\alpha (k-k ')]+\cos [\alpha (k+k ')]\}(k+k')} ,
\end{equation}
where we have used the fact that only even integrands can survive. This can be seen by performing a
coordinate transformation in $k$ space,
\begin{eqnarray}
\label{nulask}
u\dee\frac{1}{\sqrt{2}}(k+k') &\qquad  & k\dee\frac{1}{\sqrt{2}} \left(u+w\right)\nonumber\\
w\dee\frac{1}{\sqrt{2}}(k-k') & \qquad & k'\dee\frac{1}{\sqrt{2}} \left(u-w\right) ,
\end{eqnarray}
changing the integral domains accordingly, $\int_{0}^{\infty} du  \int_{-u}^{u} dw$, and noting
that $\Psi (u+w) \Psi (u-w)$ is even under the change $w\to -w$. Note that now Eq.~(\ref{guitotalpr}) is anti-symmetric under the change $\phi \to -\phi$ and again we have that $d\alpha / d\phi =0$ but at $\phi =0$. Hence, the Bohmian trajectories must certainly present a bounce when it crosses the line $\phi =0$, and
if they start at $v_r\to\infty$ in a classical contraction from infinity, they must necessarily end at $v_l\to \infty$ in
classical expansion to infinity, realizing a bounce at $\phi=0$ and never reaching the singularity, in a symmetric trajectory in $\phi$.
On the other hand,
if they start at the singularity in $v_l\to -\infty$, they come back to the singularity at $v_r\to -\infty$, with the turning
point taking place at $\phi=0$. Again, contrary to the consistent history conclusion, any universe history as described by
these Bohmian trajectories coming from infinity must go back to infinity, and any Bohmian trajectory coming from the singularity
must go back to the singularity.

Note that the line $\alpha =0$, where these non-classical behaviors are strong, corresponds, in our units, to $a_{\rm phys} = l_{\rm pl}$, where $l_{\rm pl}$ is the Planck length. All these features can be seen numerically with a particular example. Let us take, for instance,
\begin{equation}
\label{psipr}
\Psi (k) = \ e^{-(|k| - d)^2/\sigma ^2} ,
\end{equation}
with $\sigma << 1$ and $d\geq 1$. This is a real and even $\Psi (k)$, consisting of two sharply
peaked gaussians centered at $\pm d$. The wave function reads

\begin{eqnarray}
\label{solf+2}
\Psi(v_l,v_r) &=& \int_{0}^{\infty} d k \Psi(k)\ e^{ikv_l} + \int_{-\infty}^{0} d k \Psi(k)\ e^{ikv_r} \nonumber \\
&\approx& \int_{-\infty}^{\infty} d k \ e^{-(k - d)^2/\sigma ^2}\ e^{ikv_l} +
\int_{-\infty}^{\infty} d k \ e^{-(k + d)^2/\sigma ^2}\ e^{ikv_r}\quad.
\end{eqnarray}

\end{widetext}

\begin{figure}
\fbox{
\includegraphics[height=80mm,width=80mm,angle=270]{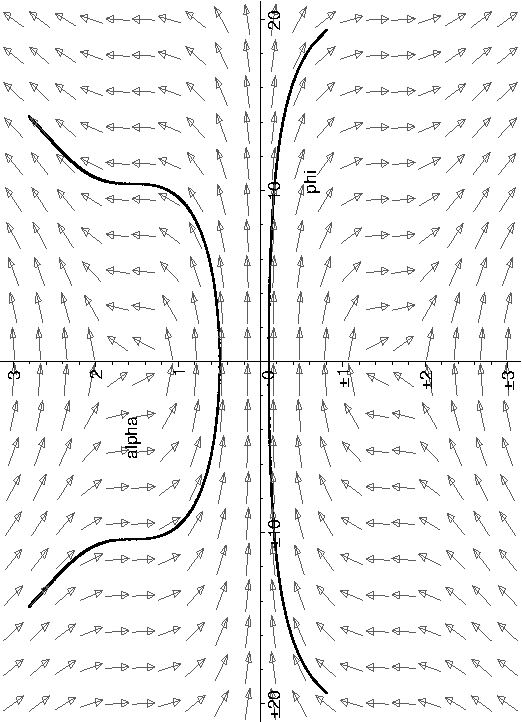}}
\caption{The field plot shows the family of trajectories for the planar system given by (\ref{guialpha}) (\ref{guiphi}) for the wave functional (\ref{solf+2}). Two of them that describe their general behavior are depicted in solid line: the first representing a bouncing universe while the second one corresponds to a universe which begins and ends in singular states (``big bang - big crunch'' universe). }
\label{traj}
\end{figure}

The Bohmian trajectories associated with this wave functions can be seen from figure 1. We can distinguish two kind of trajectories. The upper half of the figure contains trajectories describing bouncing universes while the lower half corresponds to universes that begins and ends in singular states (``big bang - big crunch'' universe).

In general, there is also the possibility of trajectories describing cyclic universes. In Ref.~\cite{nelson1}, it was considered Bohmian trajectories associated with wave functions similar to the above one, but without the restriction to positive frequencies only. Considering both positive and negative frequencies, there are oscillatory trajectories in $\phi$. In this case, if one wishes to interpret
$\phi$ as time, this corresponds to creation and annihilation of expanding and contracting universes that exist for a very short duration. This fact suggests that one cannot understand Eq.~(\ref{hoqg2p}) as a continuity equation for an ensemble of trajectories with a distribution of initial conditions given by $R^2$ in this Bohmian approach with guidance relations defined as in Eqs. (\ref{guialpha}) and (\ref{guiphi}).
In fact, this interpretation of a continuity equation would be possible only if Eq.~(\ref{hoqg2p}) could be reduced to the form
\begin{equation}
\label{cont}
\frac{\partial R^2}{\partial\phi}+\frac{\partial}{\partial\alpha}\biggl(R^2\frac{d\alpha}{d\phi}\biggr) = 0 ,
\end{equation}
with $d\alpha/d\phi$ given by Eq.~(\ref{guitotal}),

\begin{equation}
\label{fluidvelocity}
\frac{d\alpha}{d\phi} = -\frac{\partial S/\partial \alpha}{\partial S /\partial \phi} .
\end{equation}
It can be shown, using Eqs.~(\ref{hoqgp},\ref{hoqg2p},\ref{guialpha},\ref{guiphi}), that this is
possible if and only if

\begin{equation}
\frac{\partial S}{\partial \alpha}\frac{\partial^2 S}{\partial\alpha\partial\phi}=
\frac{\partial S}{\partial \phi}\frac{\partial^2 S}{\partial\phi\partial\phi} ,
\end{equation}
which implies that $\ddot{\phi}=0$, stating that $\phi$ is a monotonic function
of coordinate time. This is a strong restrictive condition, which cannot be satisfied
by general quantum states as the ones discussed above. In general, $\ddot{\phi}\neq 0$.
Hence, Eq.~(\ref{hoqg2p}) cannot be interpreted as a continuity equation in $\phi$ time for the ensemble of trajectories given by
Eq.~(\ref{fluidvelocity}) with distribution $R^2$, even in the single frequency approach
where one has a Schr\"odinger-like equation.

If, however, one insists in interpreting $\phi$ as the time variable, then
one would have to face the situation of creation and annihilation of universes which is a typical
feature of relativistic quantum theory. Accordingly, the lost of a continuity equation for $R^2$ can be associated with the non-conservation of the number of trajectories of this ensemble in the $\phi$ time.

Normally the de~Broglie-Bohm theory of a Schr\"odinger-like equation
furnishes, besides the quantum trajectories, a probabilistic measure for these trajectories.
This is not the case here since the kinetic term in the present Schr\"odinger-like equation
is not canonical, hence, it is not of the form $g_{ij}(x) p^i p^j$, where $g_{ij}(x)$ has an euclidean
signature.

Concluding this section, in the consistent histories approach we may have the notion
of probabilities but we are not allowed to investigate non-classical properties of the
universe in any finite $\phi$ time, or to have more than two snapshots of any non-classical
universe.

On the contrary, in the de~Broglie-Bohm theory we can investigate the entire evolution of the universe, but we loose the notion of probability.

It is worth remarking that, in addition to the usual debate related to different approaches to describe a quantum system, every quantum model of the universe has to face a non-trivial problem. Quantum cosmology deals with a single system, which forbids us to repeat experiments, hence posing peculiar issues associated with the physical meaning of any kind of probability in this context. Thus, the lack of probabilistic predictions in quantum cosmology should not be taken a priori as a deficiency of the formalism. On the contrary, one should carefully analyze if one can consistently extract information and predictions from the model without a notion of probability.

We should stress that one can recover the probabilistic predictions in quantum cosmology using the de~Broglie-Bohm theory when one implements a more
complex modeling of the universe, adding new degrees of freedom (here, we have only one degree of freedom). In that case, a probability measure naturally appears in the quantum description of the sub-systems of the universe (see Ref.~\cite{falciano09} for details), and the usual Born rule
can be recovered. In that case, for the sub-systems, the consistent histories and the de Broglie-Bohm approaches should coincide. 

\section{The Klein-Gordon Approach}\label{KGapp}

In the Wheeler-De Witt equation for a free massless scalar field, one can define its square-root and construct a Schr\"odinger-like equation as we have discussed in sections \ref{CSsec} and \ref{3hist}. However, there are other quantization schemes where the restriction to a single frequency sector is not necessary. A promising alternative approach to quantum cosmology using the consistent histories quantization is to consider the full Klein-Gordon equation. In this approach both energy sectors, positive and negative, are simultaneously taken into account but the Hilbert space is defined with a different inner product (see \cite{halli2} and references therein).

Following closely Ref.~\cite{halli}, one can define the  eigenstates associated to the position operator as
\begin{eqnarray}
\ket{x}
&=& \frac{1}{\sqrt{2\pi}} \int_{-\infty}^{\infty} \frac{dk}{2|k|} \ e^{i|k |\phi -ik\alpha} \ket{k_{+}}\nonumber\\
&&+ \frac{1}{\sqrt{2\pi}} \int_{-\infty}^{\infty} \frac{dk}{2|k|} \ e^{-i|k |\phi -ik\alpha}\ket{k_{-}}\nonumber\\
&=&  \ket{x_{+}} + \ket{x_{-}}  \label{position}\quad,
\end{eqnarray}
where $\ket{k_{\pm}}$ are eigenstates of the $\hat{k}$ operator such that $\hat{k}\ket{k_{\pm}} = k \ket{k_{\pm}}$ and $\hat{k}_0\ket{k_{\pm}} = \pm |k|\ \ket{k_{\pm}}$. Note that the position eigenstates are not orthogonal, \ie
\[
\bra{x}x'\ra=G^{(+)}(x,x')+G^{(-)}(x,x')
\]
where
\begin{equation}\label{Gpm}
G^{(\pm)}(x,x')\dee \frac{1}{2\pi} \int_{-\infty}^{\infty} \frac{dk}{2|k|} \ e^{\mp i |k|(\phi-\phi') \pm i k(\alpha-\alpha')}\quad ,
\end{equation}
are respectively the positive and negative Wightman functions. The positive and negative position eigenstates satisfy a completeness relation that reads
\[
\1 = i\int d\alpha \ \Big(
\ket{x_{+}}
\ov{\partial}_\phi
\bra{x_{+}}-
\ket{x_{-}}\ov{\partial_{\phi}}\bra{x_{-}}
\Big)\quad.
\]

Given these position eigenstates we can define the induced Klein-Gordon inner product as

\begin{equation}
\label{inner2}
(\Psi,\Phi)\dee i\int d\alpha \ \Big(\Psi_{+}^{*} \ov{\partial_{\phi}} \Phi_+ -
\Psi_{-}^{*} \ov{\partial_{\phi}} \Phi_-\Big)
\end{equation}
where $\Psi_{\pm}(\alpha,\phi)$ denotes the positive (negative) frequency solutions of the Klein-Gordon equation which are given by the projection of the wave function in the position eigenstates. Recalling that $v_l = \alpha + \phi$ and $v_r = \alpha - \phi$, we have
\begin{align}
\label{wavefunction+}
\Psi_{+} (\phi,\alpha) &= \bra{x_{+}} \Psi\ra \nonumber\\
=&\frac{1}{\sqrt{2\pi}}
\biggl[\int_{0}^{\infty}{dk} \ e^{i k v_r} \Psi_{+} (k)+
\int_{-\infty}^{0} dk \ e^{i k v_l} \Psi_{+} (k)\biggr] \nonumber\\
=&\Psi_{+}^r (v_r) + \Psi_{+}^l (v_l)
\end{align}
and
\begin{align}
\label{wavefunction-}
\Psi_{-} (\phi,\alpha) &= \bra{x_{-}} \Psi\ra \nonumber\\
=&\frac{1}{\sqrt{2\pi}}
\biggl[\int_{0}^{\infty} dk \ e^{i k v_l} \Psi_{-} (k)+
\int_{-\infty}^{0} dk \ e^{i k v_r} \Psi_{-} (k)\biggr] \nonumber \\
=&\Psi_{+}^r (v_r) + \Psi_{+}^l (v_l)
\end{align}
with
\begin{equation}
\label{wavefunctionk}
\Psi_{\pm} (k) = \frac{\bra{k_{\pm}} \Psi\ra }{2|k|}\quad .
\end{equation}

One of the key features of this inner product is that for an arbitrary wave function the quantity
\begin{eqnarray}
\label{posinner}
(\Psi,\Psi)&=&
i\int d\alpha \ \Big(\Psi_{+}^{*} \ov{\partial_{\phi}} \Psi_+ -
\Psi_{-}^{*} \ov{\partial_{\phi}} \Psi_-\Big)\nonumber\\
&=&2\int_{-\infty}^{\infty} dk |k| \Big(\big|\Psi_{+}{(k)}\big|^2 + \big|\Psi_{-}{(k)}\big|^2\Big)\quad ,\quad
\end{eqnarray}
is positive definite.

Once again we shall be interested in calculating the probability of the universe in a given time ($\phi$) to have a size within the range $\Delta$ or in its complement $\bar{\Delta}$. For a given initial state $\Psi (\phi,\alpha)$, we can construct the decoherence functional for a set of histories as proposed in Ref.~\cite{halli} and then take the limit of infinite past $\phi_1 \to -\infty$ and infinite future $\phi_2\to +\infty$.

\begin{widetext}
The off-diagonal terms of decoherence between histories that cross the surface $\phi=$const. within region $\Delta$ or in $\bar{\Delta}$ is given by
\begin{align}
\label{decoherence}
&D(\Delta,\bar{\Delta})=\int_{\Delta}d\alpha\int_{\bar{\Delta}} d\alpha '
\Big[\Psi_{+}^{*}(\alpha ',\phi ') \ov{\partial_{\phi '}} G^{(+)}(\alpha ',\phi ';\alpha ,\phi)
\ov{\partial_{\phi }}\Psi_+(\alpha,\phi) +\Psi_{-}^{*}(\alpha ',\phi ') \ov{\partial_{\phi '}} G^{(-)}(\alpha ',\phi ';\alpha ,\phi)
\ov{\partial_{\phi }}\Psi_-(\alpha,\phi)\Big]\quad.
\end{align}

Omitting the negative frequency terms and defining the region $\Delta=(-\infty, \alpha *)$ we have
\begin{eqnarray}
\label{deco2}
D(\Delta,\bar{\Delta})&=& \int_{-\infty}^{\alpha *}d\alpha\int_{\alpha *}^{\infty} d\alpha '
\bigg[\Psi_{+}^{*}(\alpha ',\phi ') \partial_{\phi '} G^{(+)}(\alpha ',\phi ';\alpha ,\phi)
\partial_{\phi }\Psi_+(\alpha,\phi) +
\Psi_+(\alpha,\phi)\partial_{\phi} G^{(+)}(\alpha ',\phi ';\alpha ,\phi)
\partial_{\phi '}\Psi_{+}^{*}(\alpha ',\phi ')-\nonumber\\
&&\Psi_{+}^{*}(\alpha ',\phi ') \Psi_+(\alpha,\phi)
\partial_{\phi} \partial_{\phi '} G^{(+)}(\alpha ',\phi ';\alpha ,\phi)
+ G^{(+)}(\alpha ',\phi ';\alpha ,\phi) \partial_{\phi '}\Psi_{+}^{*}(\alpha ',\phi ')\partial_{\phi}\Psi_+(\alpha,\phi)\bigg]\quad,
\end{eqnarray}
where in our specific case, the Green functions $G^{(\pm)}$ eq.~\eqref{Gpm} read
\begin{equation}
\label{green}
G^{(\pm)} = \frac{1}{2\pi} \biggl\{\int_{0}^{\infty} \frac{dk}{2|k|} e^{\pm i k(v_r-v_r ')}
+ \int_{-\infty}^{0} \frac{dk}{2|k|} \ e^{\pm  i k(v_l-v_l ')}\biggr\}.
\end{equation}

Let us evaluate Eq.~(\ref{deco2}) term by term. The first two terms involving first derivatives of
the Green function are zero as long as, when $\phi ' = \phi$, $\partial_{\phi} G^{(+)}(\alpha ',\phi ';\alpha ,\phi)$ yields a $\delta (\alpha ' - \alpha)$ and the integrations are in disjoint
domains of $\alpha$. The third and fourth terms are more involved and require more attention. Each of them involves eight terms, where four of them are null. These are the terms involving either $v_r$ or $v_l$ only. For instance, the term proportional to
\begin{eqnarray}
&&\int_{-\infty}^{\alpha *}d\alpha\int_{\alpha *}^{\infty} d\alpha '
\int_{0}^{\infty} dk'  \ e^{-i k ' v_r '} \Psi_{+}^{*} (k')
\int_{0}^{\infty} dk \ e^{i k v_r} \Psi_{+} (k)
\int_{0}^{\infty} dk '' |k ''| \ e^{i k ''(v_r '-v_r)} \nonumber \\
&=& \int_{0}^{\infty} dk '' |k ''|\int_{-\infty}^{\alpha * - \phi} d v_r \ e^{i (k - k '') v_r}
\int_{\alpha * - \phi}^{\infty} d v_r ' \ e^{i (k '' - k ') v_r '}
\int_{0}^{\infty} dk' \Psi_{+}^{*} (k')
\int_{0}^{\infty} dk \Psi_{+} (k)
\end{eqnarray}
is zero in the limit $\phi \to \pm\infty$. After the change of variables $d\alpha \rightarrow d v_r$, it remains a $\phi$ dependence only in the limits of integration, hence we can safely take the limit $\phi \to \pm\infty$ before integrating the expression which makes it to go to zero. Notwithstanding, the mixed terms that include both $v_l$ and $v_r$ have a complete different structure. These terms are proportional to
\begin{eqnarray}
&&\int_{-\infty}^{\alpha *}d\alpha\int_{\alpha *}^{\infty} d\alpha ' \biggl[ \nonumber \\
&-&\int_{0}^{\infty} dk' \ e^{-i k ' v_r '} \Psi_{+}^{*} (k')
\int_{-\infty}^0 dk\ e^{ i k v_l }\Psi_{+} (k)\biggl(
\int_{0}^{\infty} dk '' k ''\ e^{ i k ''(\alpha '-\alpha)}
-\int_{-\infty}^0 dk '' k ''\ e^{ i k ''(\alpha '-\alpha)} \biggr)\nonumber \\
&-& \int_{-\infty}^0 dk' \ e^{ -i k ' v_l '} \Psi_{+}^{*} (k')
\int_0^{\infty} dk\ e^{ i k v_r} \Psi_{+} (k)\biggl(
\int_{0}^{\infty} dk '' k ''\ e^{ i k ''(\alpha '-\alpha)}
-\int_{-\infty}^0 dk '' k ''\ e^{ i k ''(\alpha '-\alpha)} \biggr)\nonumber \\
&+& \int_{0}^{\infty} dk' k'\ e^{ -i k ' v_r '} \Psi_{+}^{*} (k')
\int_{-\infty}^0 dk k\ e^{ i k v_l} \Psi_{+} (k)\biggl(
\int_{0}^{\infty} \frac{dk ''}{k ''}\ e^{ i k ''(\alpha '-\alpha)}
-\int_{-\infty}^0 \frac{dk ''}{k ''}\ e^{ i k ''(\alpha '-\alpha)} \biggr)\nonumber \\
&+& \int_{-\infty}^0 dk' k'\ e^{ -i k ' v_l '} \Psi_{+}^{*} (k')
\int_0^{\infty} dk k\ e^{ i k v_r} \Psi_{+} (k)\biggl(
\int_{0}^{\infty} \frac{dk ''}{k ''}\ e^{ i k ''(\alpha '-\alpha)}
-\int_{-\infty}^0 \frac{dk ''}{k ''}\ e^{ i k ''(\alpha '-\alpha)} \biggr)\biggr]\quad
\end{eqnarray}

Let us take, for instance, the first term of the first line of the above equation. After some change of variables we obtain

\begin{equation}
\int_{0}^{\infty} dk '' k ''
\int_{0}^{\infty} d v_l \ e^{-i v_l (k-k'')}\int_{0}^{\infty}dv_r ' \ e^{i v_r ' (k''-k')}
\int_{0}^{\infty} dk' \Psi_{+}^{*} (k')
\int_{-\infty}^0 dk \Psi_{+} (k) \ e^{i \alpha * (k-k')}\ e^{i \phi (k+k')}\quad .
\end{equation}
\end{widetext}
Using again that
\begin{equation}
\int_0^\infty d\alpha\; e^{-i\alpha x}= \pi \delta(x) -i\text{p.v.}\left(\frac{1}{x}\right)\quad ,
\end{equation}
one gets four integrals. One of these integral ends up as
\begin{equation}
\int_{0}^{\infty} dk '' \frac{k ''}{(k''-k')(k''-k)}\quad ,
\end{equation}
which has an ultra-violet logarithmic divergence at infinity. Using the same reasoning in the first term of the third line, one gets the integral
\begin{equation}
\int_{0}^{\infty} \frac{dk ''}{k''(k''-k')(k''-k)}\quad,
\end{equation}
which now presents an infra-red logarithmic divergence the origin. The crucial point is that these are different divergencies which cannot cancel each other out.

In this way, the decoherence functional cannot be made diagonal, and hence we cannot construct consistent histories. 

In fact, one could have anticipated this result. Note that the Wheeler-DeWitt equation we are considering is completely analogous to the Klein-Gordon equation for a massless relativistic particle. However, 
as pointed out in Ref.~\cite{halli}, where the decoherence functional was constructed for a massive relativistic particle, it was observed that the off-diagonal terms $D(\Delta,\bar{\Delta})$ may become negligible only if the region $\Delta$ and its complement are much larger than the Compton wavelength $m^{-1}$ of the particle. If we naively take the limit $m\rightarrow0$, there is no region $\Delta$ in which the off-diagonal terms can become negligible. Note, however, that the $m\rightarrow0$ limit of a Klein-Gordon particle is tricky and subtle. That is why we have constructed the decoherence functional for the equivalent of a massless scalar field from the beginning. 

Note, however, that if one applies the de~Broglie-Bohm quantum theory to the same problem, one can obtain information about the behavior of the early universe, and Bohmian bouncing trajectories appear in many circumstances. This was done in detail in Ref.~\cite{nelson1}.

\section{Conclusion}

We have shown that claims asserting that the Wheeler-DeWitt quantization does not eliminate the classical cosmological singularity are not correct without specifying the quantum interpretation one is adopting and the quantization procedure one is taking.

In fact, there are several papers showing quantum bounces in the Wheeler-DeWitt quantization scenario that were published much before this sort of claims have been first presented (a few examples of the long list of papers are \cite{nelson1,nelson2}).

This, however, does not diminish in any way the importance of the series of results obtained in the context of loop quantum cosmology. Indeed, loop quantum cosmology has an important advantage over Wheeler-DeWitt quantum cosmology inasmuch it has a strong connection with loop quantum gravity. Loop quantum gravity has much less conceptual problems as a quantum theory of gravity than the canonical quantization procedure that leads to the Wheeler-DeWitt equation. Therefore, the existence of bouncing solutions in loop quantum cosmology is a significantly relevant result. Note, however, that people working in loop or non-loop quantum cosmology must be very precise on what quantum theory they are taking to interpret their results.

In connection with the conclusion of Ref.~\cite{CS}, we have shown here that the answers given by the consistent histories approach are quite fragile. In fact, the existence of a quantum bounce strongly depends on the family of histories one is taking. One can argue that the family with only two moments of time, where quantum bounces do not exist, encompass the families of histories with three moments of time. Take, however, a genuine quantum state. In the two-time family we are sure that there is no quantum bounce, but in the three-time family this question cannot even be posed. This is characteristic of the consistent histories approach: the notion of truth depends on the family of histories one is taking. This ambiguity on the notion of a true statement in the consistency histories approach can be made quite dramatic in other circumstances \cite{ghirardi,ghirardi2}.

Finally, we would like to stress that the results of this paper go much beyond the question about the existence of a quantum bounce. It shows that different quantum theories may present quite discrepant results when this system is the Universe. This finding points out to a hope that maybe in cosmology one can find a way do discriminate between the many proposed quantum theories which, asides subjective and philosophical preferences, have all the same scientific status in the laboratory.

\begin{acknowledgments}
R.P. was funded by a PCI scholarship from MCTI/CNPq of Brazil. NPN would like to thank CNPq of Brazil for financial support. ESS would like to thank CNEN and CBPF-MCTI for technical support. We also would like to thank `Pequeno Seminario' of CBPF's Cosmology Group
for useful discussions.
\end{acknowledgments}

\appendix

\section{Distributions}\label{appdist}

We use throughout the text the following distribution(see for instance \cite{dist,feynman}):

\begin{equation}
d(x) \dee \int_0^\infty d\alpha\; e^{-i\alpha x}= \pi \delta(x) -i\text{p.v.}\left(\frac{1}{x}\right).
\end{equation}

It is the Fourier transform of a Heaviside function. $\text{pv}\left(\frac{1}{x}\right)$ stands for principal value of $\frac{1}{x}$. It solves the following equation in the sense of distributions
\begin{equation}\label{disteq}
\text{p.v.}\left(\frac{1}{x}\right) x = 1.
\end{equation}
The action of the principal value on any function $f\in C^\infty_0(\setR)$ is given explicitly by:
\bea
\text{p.v.}\left(\frac{1}{x}\right) (f) &=& \lim_{\epsilon\rightarrow 0} \left(\int_{-\infty}^{-\epsilon}\frac{1}{x}f(x)dx + \int_\epsilon^\infty \frac{1}{x}f(x) dx\right) = \no
&=&  \int_\epsilon^\infty \frac{f(x) - f(-x)}{x} dx.
\ena

\end{document}